\documentclass[journal=jctcce,manuscript=article]{achemso}

\usepackage[version=3]{mhchem} 
\usepackage{xcolor}
\usepackage{soul}
\usepackage{siunitx}
\usepackage{gensymb}
\usepackage{float}
\usepackage{hyperref}

\author{Ofir Blumer}
\affiliation[TAU]
{School of Chemistry, Tel Aviv University, Tel Aviv 6997801, Israel.}
\author{Shlomi Reuveni}
\affiliation[TAU]
{School of Chemistry, Tel Aviv University, Tel Aviv 6997801, Israel.}
\alsoaffiliation[Sackler]
{The Center for Computational Molecular and Materials
Science, Tel Aviv University, Tel Aviv 6997801, Israel.}
\alsoaffiliation[BioSoft]
{The Center for Physics and Chemistry of Living Systems, Tel Aviv University, Tel Aviv 6997801, Israel.}
\author{Barak Hirshberg}
\email{hirshb@tauex.tau.ac.il}
\affiliation[TAU]
{School of Chemistry, Tel Aviv University, Tel Aviv 6997801, Israel.}
\alsoaffiliation[Sackler]
{The Center for Computational Molecular and Materials
Science, Tel Aviv University, Tel Aviv 6997801, Israel.}
\alsoaffiliation[BioSoft]
{The Center for Physics and Chemistry of Living Systems, Tel Aviv University, Tel Aviv 6997801, Israel.}

\title[]{\vspace{-1cm}Short-Time Infrequent Metadynamics for Improved Kinetics Inference}


\begin{document}

\begin{abstract}

Infrequent Metadynamics is a popular method to obtain the rates of long timescale processes from accelerated simulations. 
The inference procedure is based on rescaling the first-passage times of Metadynamics trajectories using a bias-dependent acceleration factor.  
While useful in many cases, it is limited to Poisson kinetics, and a reliable estimation of the unbiased rate requires slow bias deposition and prior knowledge of efficient collective variables.
Here, we propose an improved inference scheme, which is based on two key observations:
1) The time-independent rate of Poisson processes can be estimated using short trajectories only.
2) Short trajectories experience minimal bias, and their rescaled first-passage times follow the unbiased distribution even for relatively high deposition rates and suboptimal collective variables.
Therefore, by limiting the inference procedure to short timescales, we obtain an improved tradeoff between speedup and accuracy at no additional computational cost, especially when employing suboptimal collective variables.
We demonstrate the improved inference scheme for a model system and two molecular systems.

\end{abstract}




\section*{Introduction}

Molecular dynamics (MD) simulations are widely used to study complex systems at the microscopic level. Their atomic resolution allows evaluating thermodynamic and kinetic properties, but also limits the accessible timescales~\cite{tiwary_metadynamics_2013,salvalaglio_assessing_2014,valsson_enhancing_2016,palacio-rodriguez_transition_2022}. Therefore, long timescale processes, such as protein folding or crystal nucleation, are almost never studied using brute-force, long simulations~\cite{ray_kinetics_2023}. Instead, enhanced sampling methods are usually employed. 

Various methods were developed to study long timescale processes through MD simulations.
Some use a series of short simulations to sample the unbiased kinetics, such as milestoning~\cite{faradjian_computing_2004,elber_milestoning_2020}, Markov state models~\cite{bowman_introduction_2014,husic_markov_2018}, stochastic resetting~\cite{blumer_stochastic_2022,blumer2024combining} and many others. Another approach is to introduce an external bias potential, enhancing the sampling along a low dimensional collective variables (CVs) space. The chosen CVs are usually slow modes that can distinguish between metastable states~\cite{tiwary_metadynamics_2013,ray_kinetics_2023,palacio-rodriguez_transition_2022,salvalaglio_assessing_2014,valsson_enhancing_2016}. Methods following this approach include umbrella sampling~\cite{torrie_nonphysical_1977,kastner_umbrella_2011}, conformational flooding~\cite{grubmuller_predicting_1995}, adiabatic free energy dynamics~\cite{rosso_adiabatic_2002,rosso_use_2002,abrams_efficient_2008}, On-the-fly Probability Enhanced Sampling (OPES)~\cite{invernizzi_rethinking_2020,invernizzi_exploration_2022},  Metadynamics (MetaD)~\cite{barducci_metadynamics_2011,sutto_new_2012,valsson_enhancing_2016} and more. 

Here, we focus on infrequent MetaD (iMetaD), a method to extract the unbiased kinetics from accelerated MetaD simulations. In iMetaD, several biased trajectories are initiated and stopped after a first-passage criterion is fulfilled. The first-passage time (FPT) of each trajectory is then rescaled by an acceleration factor that depends on the external bias deposited along the trajectory~\cite{tiwary_metadynamics_2013,valsson_enhancing_2016}.
The method assumes that the underlying process obeys Poisson kinetics, and the unbiased rate is estimated by fitting the rescaled FPTs to an exponential distribution~\cite {salvalaglio_assessing_2014}. 

The key assumption of iMetaD is that no bias is deposited near the transition state~\cite{tiwary_metadynamics_2013, ray_kinetics_2023}. This assumption fails for high bias deposition rates or suboptimal CVs, that lead to hysteresis and bias over-deposition~\cite{salvalaglio_assessing_2014,ray_kinetics_2023,ray_rare_2022}. 
Unfortunately, finding good CVs in complex systems remains a great challenge~\cite{invernizzi_making_2019,doi:10.1021/acs.jctc.3c00448}, despite recent developments~\cite{mendels_folding_2018,mendels_collective_2018,bonati_deep_2021,sidky_machine_2020,chen_collective_2021,liu_graphvampnets_2023,doi:10.1021/acs.jctc.3c00051}. Thus, to improve the inference, one is usually forced to limit the bias deposition rate but this also reduces the acceleration, resulting in a tradeoff between speedup and accuracy~\cite{ray_rare_2022,blumer2024combining,palacio-rodriguez_transition_2022}.

The reliability of iMetaD is usually assessed through a procedure suggested by Salvalagio et. al.~\cite{salvalaglio_assessing_2014,ray_kinetics_2023}
A Kolmogorov–Smirnov
(KS) test~\cite{massey_kolmogorov-smirnov_1951} is performed to accept or reject the hypothesis that the rescaled FPTs are taken from an exponential distribution.
The results are considered reliable for a p-value greater than $0.05$, and the unbiased mean FPT (MFPT) is then estimated as the mean of the exponential fit to the rescaled FPT distribution.
Trajectories with hysteresis or over-deposition contribute unrealistically long rescaled FPTs, leading to distributions that are broader than exponential and failure of the KS test. 
In this paper, we propose an improved inference scheme which deals with this prevalent problem, extending the range of applicability of iMetaD. We also compare our method with the Kramers Time-Dependent Rate (KTR) method that was recently introduced with a similar goal in mind~\cite{palacio-rodriguez_transition_2022,ray_kinetics_2023}.

The scheme relies on two key observations: 1) Since exponential distributions are characterized by a single parameter -- their time-independent kinetic rate -- short simulations, showing a single transition each, are sufficient to estimate the full distribution reliably. 2) The rescaling procedure of iMetaD is more reliable for short trajectories, experiencing minimal bias. This can be seen from the rescaled FPT distribution, which often follows the unbiased distribution at short times even when using high bias deposition rates or suboptimal CVs. The improved scheme is inspired by our previous work, combining MetaD with stochastic resetting (SR)~\cite{blumer2024combining}. Previously, we showed that SR provides enriched sampling of short timescales, leading to an improved tradeoff between speedup and accuracy. Interestingly, these observations are not limited to simulations with SR. 

We next show how to exploit our observations to build an improved kinetics inference scheme for iMetaD simulations. We refer to this scheme as Short Time iMetaD (ST-iMetaD). ST-iMetaD extends the applicability of iMetaD even to high bias deposition rates and suboptimal CVs, reducing the prediction errors by orders of magnitude in comparison to the standard procedure.
We demonstrate its advantages in three systems of increasing complexity: the two-dimensional Wolfe-Quapp potential, alanine dipeptide in vacuum, and the unfolding of the chignolin mini-protein in water.

\section*{The ST-iMetaD scheme}

We present ST-iMetaD through the example of the Wolfe-Quapp potential. It is a two states model, previously used to study the performance of suboptimal CVs~\cite{invernizzi_making_2019,ray_rare_2022}. Its exact form is given in the Methods section, as well as all simulations details. We first performed 1000 brute-force, standard MD simulations, to obtain the unbiased FPT distribution, and found the MFPT to be $\sim 110 \, ns$. A KS test confirmed that the unbiased distribution is exponential (a p-value of 0.81). 

\begin{figure}[t]
  \includegraphics[width=\linewidth]{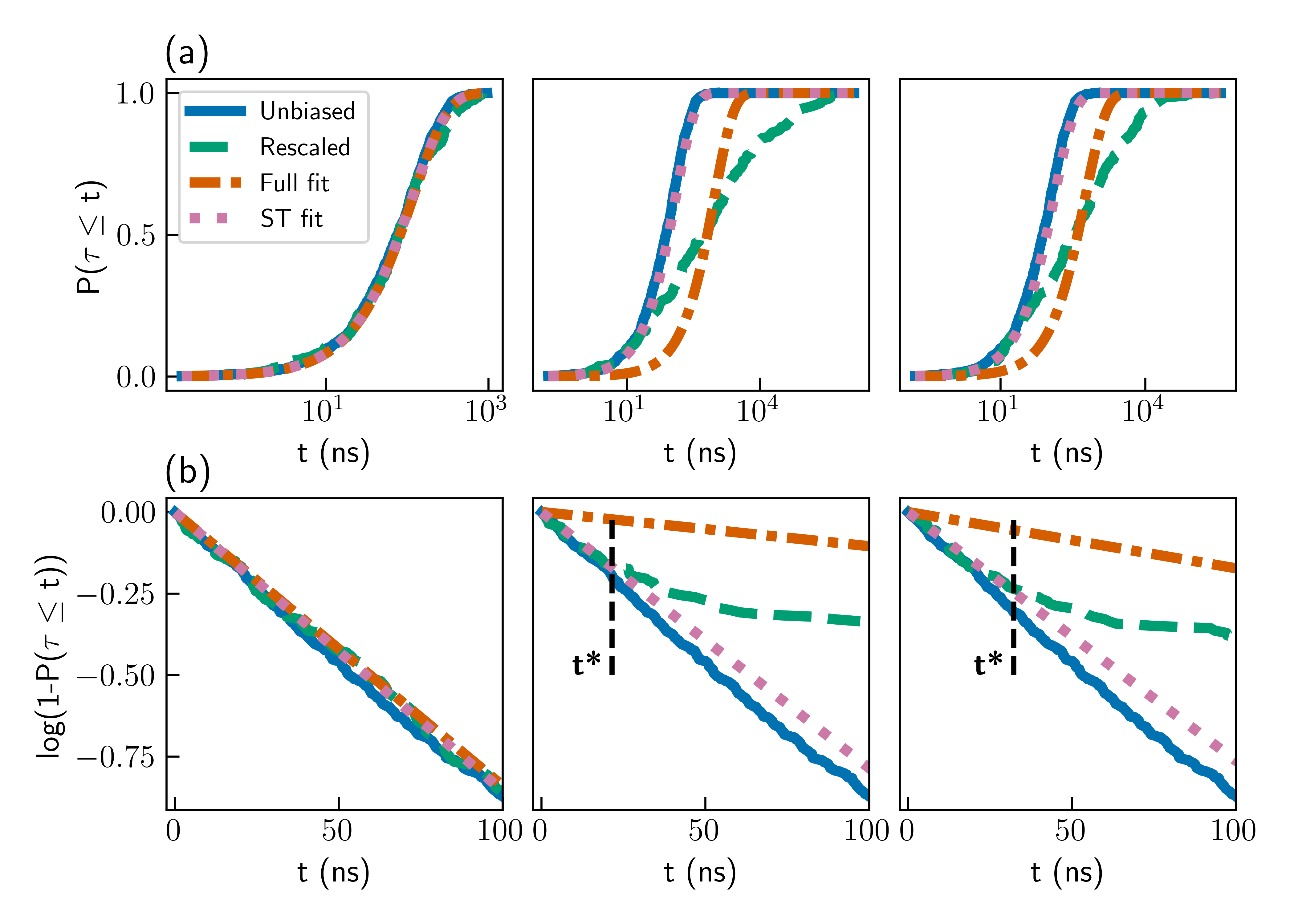}
  \caption{
  (a) CDF profiles and (b) survival functions for simulations of the Wolfe-Quapp potential. Results for unbiased FTPs (blue solid lines), rescaled FPTs (dashed green lines), exponential fits to the rescaled CDF in the entire range (orange dash-dotted lines), and linear fits to the survival functions at $t \le t^*$ (pink dotted lines).  Results are shown for iMetaD simulations using a good CV and bias deposition rate of $10 \, ns^{-1}$ (left), a good CV and bias deposition rate of $1000 \, ns^{-1}$ (middle) and a suboptimal CV and bias deposition rate of $200 \, ns^{-1}$ (right). The black dashed lines mark $t=t^{*}$.} 
  \label{fig:explain}
\end{figure}

Next, we performed 200 iMetaD simulations with a good CV and slow bias deposition rate of $10 \, ns^{-1}$, updating the bias every $10^5$ timesteps. The quality of the CV was proved using a committor analysis~\cite{bolhuis_t_2002,PETERS2017539} (see the SI for details). With this choice of parameters, we expect the underlying assumptions of iMetaD to be valid, and the original inference scheme to be accurate. Indeed, the MFPT estimated through the standared inference procedure is $119 \, ns$, in good agreement with the true value. A p-value of $0.25$ confirms the reliability of the results. The left panel of Figure \ref{fig:explain}(a) shows the cumulative distribution functions (CDF) $\mbox{P}\left(\tau \le t\right)$ for both the unbiased FPTs (blue solid line) and the rescaled FPTs (green dashed line). An exponential fit to the CDF of the rescaled FPTs is given in an orange dash-dotted line. We find a good agreement between all three curves, showing that the standard inference procedure is adequate in this case.

We then performed iMetaD simulations using the same CV but with a higher bias deposition rate of $1000 \, ns^{-1}$ (every $1000$ timesteps), which is expected to give poor inference due to hysteresis. The obtained MFPT, $953 \, ns$, overestimates the true value by an order of magnitude, and the p-value of the KS test drops to $\num{2e-11}$, indicating that the results are unreliable. The middle panel of Figure \ref{fig:explain}(a) again shows the CDF for the unbiased and rescaled FPTs, as well as the exponential fit to the rescaled CDF. In this case, we find that the rescaled CDF is not exponential and clearly deviates from the unbiased CDF. As a result, the exponential fit results in a wrong estimate of the rate and MFPT. 
Nevertheless, we find that the unbiased and rescaled CDFs are in very close agreement at short times. This is the first key observation of this work: short trajectories experience minimal bias and thus their rescaled CDF reflects the correct statistics for small FPTs even at high bias deposition rates.

A similar phenomenon is observed when employing suboptimal CVs. We select a moderate bias deposition rate of $200 \, ns^{-1}$, and intentionally reduce the quality of the CV by rotating it at an angle $\theta=56\degree$ relative to the good CV. The right panel of Figure \ref{fig:explain}(a) presents the resulting CDF profiles. Once again, the rescaled CDF is far from exponential (p-value of $\num{7e-9}$), and the MFPT is overestimated ($579 \, ns$). However, even though the rescaled CDF deviates from the unbiased CDF at long times, they match closely at short times.

When employing iMetaD, it is common practice to present the rescaled CDF profile, which is used for the KS test~\cite{salvalaglio_assessing_2014,palacio-rodriguez_transition_2022,ray_kinetics_2023}. However, for the rest of this paper, it would be more convenient to examine the survival function, $1 - \mbox{P}\left(\tau \le t\right)$, since its logarithm decays linearly for exponential distributions.  
Figure \ref{fig:explain}(b) gives $\log\left(1 - \mbox{P}\left(\tau \le t\right)\right)$ at $t \le 100 \, ns$ for the unbiased FPTs (blue solid lines) and for the rescaled FPTs (green dashed lines) of the simulations presented in Figure \ref{fig:explain}(a). The unbiased survival function decays linearly, as expected. When the assumptions of iMetaD hold (left panel), the rescaled survival function closely follows the unbiased one. When they break (middle and right panels), the rescaled survival function matches the unbiased one up to some finite time $t=t^*$, and decays slower at $t>t^*$.
As a result, the exponential fits to the rescaled data (orange dash-dotted lines) decay much slower than the unbiased curves. This explains the overestimated MFPT values. 
Note that we fit the rescaled survival function for all $t$ but only display $t \le 100 \, ns$.

We improve the inference by fitting a linear function $S(t)=-kt$ to the rescaled survival function at $t \le t^*$ only (dotted pink lines in Figure \ref{fig:explain}). We then estimate the MFPT as $k^{-1}$. Notice that we only fit a single parameter $k$, as the survival function must fulfill $\log\left(1 - \mbox{P}\left(\tau \le t=0\right)\right)=0$ due to normalization. In all three cases, we find that these short-time fits are closer to the unbiased survival function, and therefore lead to an improved estimate of the MFPT, as we show below. First, we explain how to choose an adequate value of $t^*$. 

We use the Pearson correlation coefficient $R$, which quantifies the quality of the linear fit to the survival function. Practically, we perform multiple fittings, with different choices of $t^*$, and select the fit resulting in the highest value of $R^2$. Our results show that this approach correctly identifies reasonable values of $t^*$, such that the rescaled survival functions match the unbiased ones at $t \le t^*$. Specifically, for a good CV and a low bias deposition rate, where the results are reliable, we find $t^* = 148 \, ns$. For the same CV and a high bias deposition rate, and for a poor CV and a moderate rate, we obtain lower values,  $t^* = 21 \, ns$ and $t^* = 32 \, ns$, respectively (black dashed lines in Figure \ref{fig:explain}(b)). 

To summarize our method, the main modification to the original inference scheme is that instead of fitting an exponential distribution to all of the data, as customary, we limit the analysis to short timescales. We perform a series of linear fits to the logarithm of the survival function at times $t \le t^*$, with different choices of $t^*$. The parameter $k$ of the best fit is taken as the kinetic rate, and the MFPT is estimated as $k^{-1}$. This enables accurate estimations of the MFPT, even with frequent bias deposition or a suboptimal CV. 

\section*{Results and discussion}

\subsection*{The Wolfe-Quapp potential}
\begin{figure}[t]
  \includegraphics[width=\linewidth]{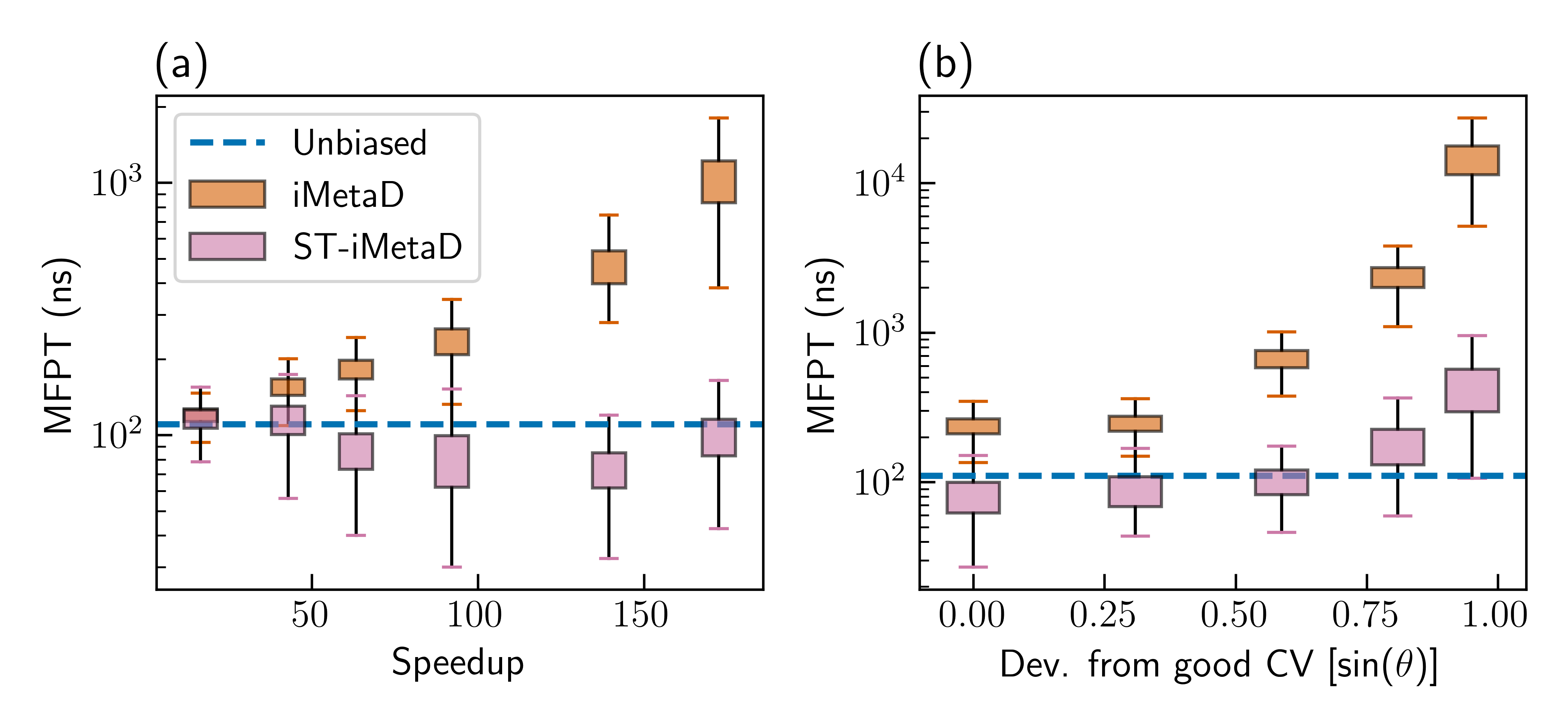}
  \caption{
  (a) Estimated MFPTs as a function of the speedup for the Wolfe-Quapp potential using a good CV and different bias deposition rates from $10$ to $1000\, ns^{-1}$. (b) Estimated MFPTs for a bias deposition rate of $200 \, ns^{-1}$, and different choices of CV. The blue lines mark the unbiased MFPT value. We employed either standard iMetaD (orange) or ST-iMetaD (pink).
  The boxes show the range between the first and third quartiles and the whiskers show extreme values within 1.5 IQR below and above these quartiles.
  }
  \label{fig:WQpotential}
\end{figure}

We first demonstrate the advantages of ST-iMetaD using the Wolfe-Quapp potential, showing that
we can use higher bias deposition rates, providing higher speedups, with minimal penalty to the inference accuracy. We define the speedup as the ratio between the unbiased MFPT and the MFPT from the biased simulations without rescaling. We ran a total of 1000 trajectories and performed a bootstrapping analysis on 1000 randomly sampled sets, each containing 200 samples. Figure \ref{fig:WQpotential}(a) shows the estimated MFPT as a function of the speedup, using a good CV and different bias deposition rates in the range of $10$ to $1000\, ns^{-1}$. The boxes show the range between the first and third quartiles (interquartile range, IQR), and the whiskers show \textit{extreme} values within 1.5 IQR below and above these quartiles. When employing standard iMetaD (orange), the estimated MFPT increases with speedup, reaching values about an order of magnitude larger than the true one at high speedups. On the other hand, when employing ST-iMetaD (pink), the estimations remain close to the true value for all speedups.

Our scheme also improves the inference from simulations performed with suboptimal CVs. For a fixed bias deposition rate of $200 \, ns^{-1}$, we gradually reduce the quality of the CV by rotating it with respect to the good CV. Figure \ref{fig:WQpotential}(b) shows that the estimated MFPT increases as the quality of the CV decreases, for both inference schemes. However, the deviation from the true value is much smaller for ST-iMetaD. The errors remain within an order of magnitude of the true value, in comparison to more than two orders of magnitude for standard iMetaD, even for very poor CVs. In the SI, we also provide a detailed comparison with the KTR method.\cite{palacio-rodriguez_transition_2022}

\begin{figure}[t]
  \includegraphics[width=\linewidth]{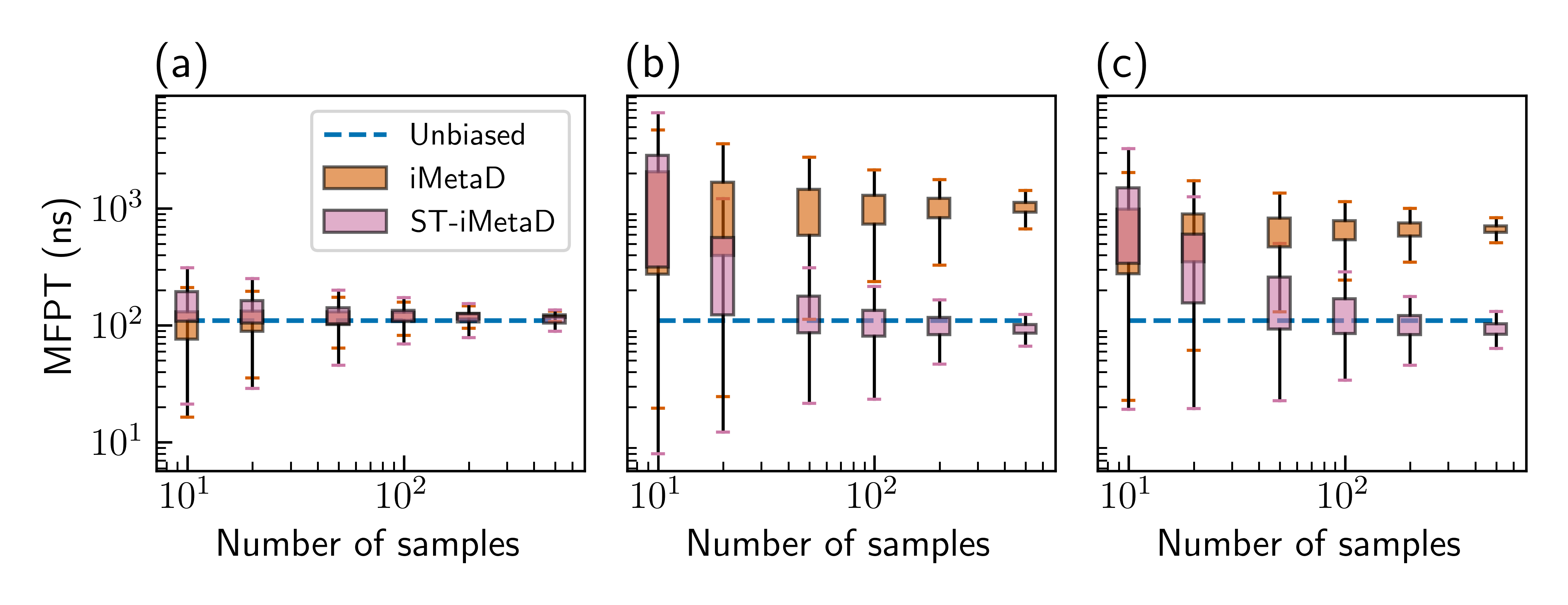}
  \caption{
  Estimated MFPT as a function of the number of sampled trajectories in each bootstrapping batch, using iMetaD (orange) or ST-iMetaD (pink), for simulations of the Wolfe-Quapp potential using (a) a good CV and bias deposition rate of $10 \, ns^{-1}$, (b) a good CV and bias deposition rate of $1000 \, ns^{-1}$, and (c) a suboptimal CV and bias deposition rate of $200 \, ns^{-1}$.
  The unbiased MFPT is given in blue dashed lines.
  The boxes show the range between the first and third quartiles and the whiskers show extreme values within 1.5 IQR below and above these quartiles.
  }
  \label{fig:sensitivity}
\end{figure}

We tested the sensitivity of ST-iMetaD to the number of sampled trajectories. For each column in Figure \ref{fig:explain}, we ran a total of 1000 trajectories and performed a bootstrapping analysis on 1000 randomly sampled sets, each containing $10,20,50,100,200,$ and $500$ samples. Figure \ref{fig:sensitivity} shows the estimated MFPT using either iMetaD (orange) or ST-iMetaD (pink) as a function of the number of samples. In Supplementary Figure S2, we also plot the dependence of $t^*$ on the batch size.
We find that iMetaD has a systematic error that is almost independent of the number of samples while both the systematic and statistical errors of ST-iMetaD diminish with additional data. With limited data of $10$ or $20$ samples, ST-iMetaD gives comparable results to iMetaD, but $50$ samples are already sufficient for a major improvement. For the remainder of the paper, we report results obtained with bootstrapping sets of 200 random samples. Equivalent figures with smaller sample sizes are provided in the SI.

\subsection*{Alanine dipeptide}

We next apply ST-iMetaD in two molecular systems,
starting with the
 well-studied example of alanine dipeptide in vacuum. Alanine dipeptide has two stable conformers, $C_{7eq}$ and $C_{7ax}$, and is usually described by two dihedral angles, $\phi$ and $\psi$, with $\phi$ serving as a good CV and $\psi$ as a suboptimal one\cite{tiwary_metadynamics_2013,salvalaglio_assessing_2014,ray_rare_2022,invernizzi_rethinking_2020}
(see Ref.\cite{tiwary_metadynamics_2013} for definitions of conformers and angles). Transitions from the $C_{7eq}$ conformer to the $C_{7ax}$ conformer have an estimated MFPT of $\sim 3.5 \, \mu s$ (see the SI for more details). We performed MetaD simulations with bias deposition rates in the range of $20$ to $1000 \, ns^{-1}$, and either the $\phi$ or $\psi$ angle as CV. Full simulation details are given in the Methods section. 

\begin{figure}[t!]
  \includegraphics[width=\linewidth]{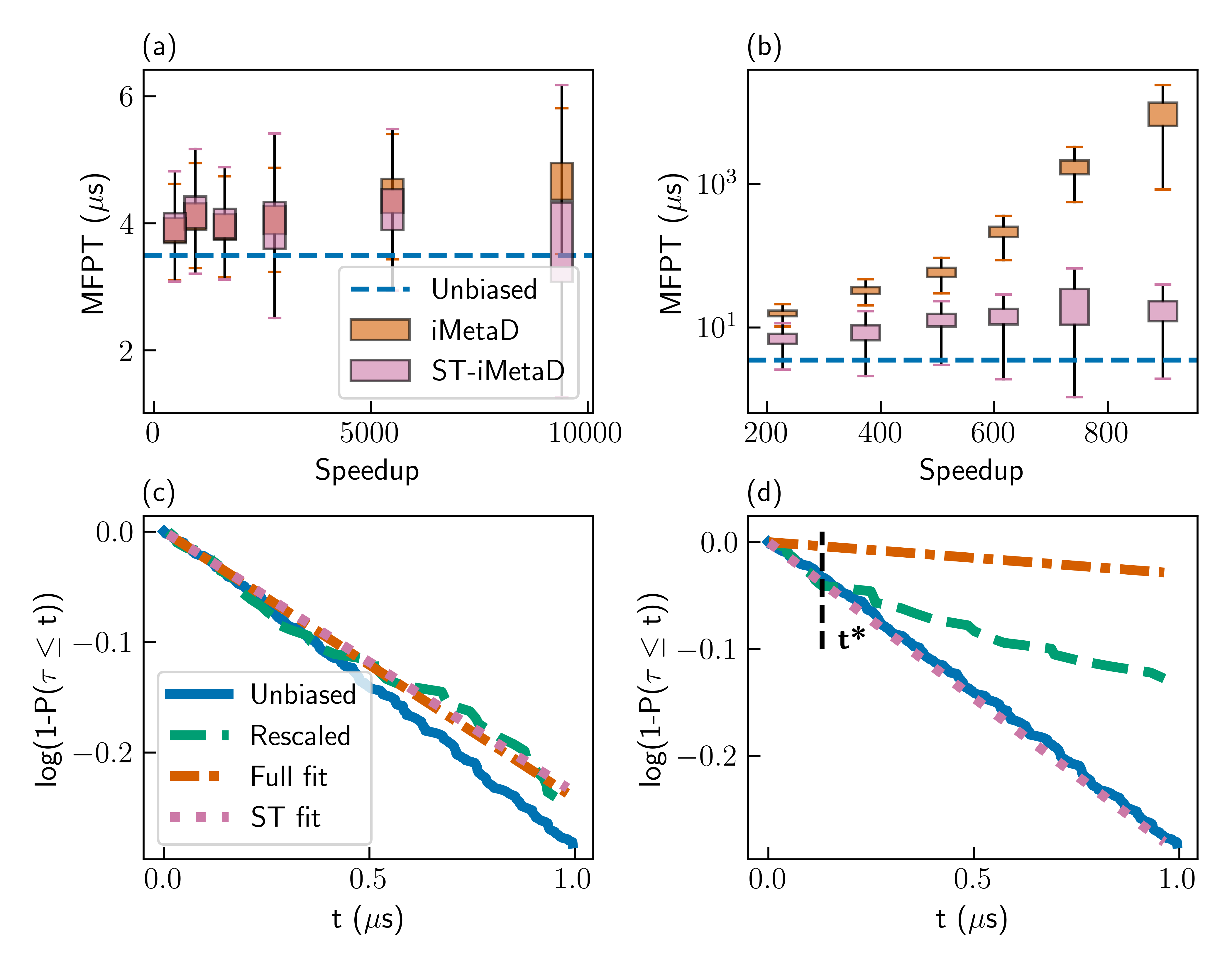}
  \caption{Upper row: Estimation of the MFPT as a function of speedup, for the $C_{7eq} - C_{7ax}$ conformer transition of alanine dipeptide in vacuum. Simulations using either (a) the $\phi$ angle or (b) the $\psi$ angle as CV, with iMetaD (orange) or ST-iMetaD (pink). The boxes show the range between the first and third quartiles and the whiskers show extreme values within 1.5 IQR below and above these quartiles. The blue dashed lines show the unbiased MFPT.
  Lower row: Survival functions $\log\left(1 - \mbox{P}\left(\tau \le t\right)\right)$ at $t \le 1 \, \mu s$ for unbiased simulations (blue solid lines) and rescaled iMetaD simulations (green dashed lines), biasing either the (c) $\phi$ angle or (d) $\psi$ angle, with bias deposition rate of $50 \, ns^{-1}$. Additional lines show exponential fits to the rescaled CDF in the entire range (orange dash-dotted lines), and linear fits to the survival functions at $t \le t^*$ (pink dotted lines). The black dashed line marks $t=t^{*}$.}
  \label{fig:ala2}
\end{figure}

Figure \ref{fig:ala2}(a) shows the estimated MFPT as a function of the speedup for simulations biasing the $\phi$ angle, through the original iMetaD scheme (orange) and ST-iMetaD (pink). The unbiased MFPT is given in a dashed blue line. The two schemes provide similar, very accurate estimations, even with frequent bias deposition. This confirms that ST-iMetaD is consistent with standard iMetaD, when the latter is reliable.

On the other hand, when $\psi$, a suboptimal CV, is employed we find a major difference between the schemes, as demonstrated in Figure \ref{fig:ala2}(b). With standard iMetaD, the estimated MFPT rapidly increases with speedup (notice the logarithmic scale), reaching errors of more than three orders of magnitude. However, with ST-iMetaD, we obtain estimations within up to about an order of magnitude from the true value, for all speedups.

We validate the underlying assumptions of ST-iMetaD by examining the survival functions. Panels (c) and (d) of Figure \ref{fig:ala2} show the survival functions for the unbiased FPT distribution (solid blue lines), the rescaled FPT distributions (dashed green lines), the fit of iMetaD (dash-dotted orange lines) and the fit of ST-iMetaD (pink dotted lines). Results are shown for simulations with a moderate bias deposition rate of $50 \, ns^{-1}$, which is standard for iMetaD~\cite{tiwary_metadynamics_2013,salvalaglio_assessing_2014}. 
Using the $\phi$ angle as CV, the rescaled survival function decays at a similar rate as the ubniased one (panel (c)). We determine $t^*=12.6 \, \mu s$ using the procedure described before, and the two fits coincide. Using the $\psi$ angle as CV, the rescaled survival function quickly deviates from the ubniased one, but is accurate at short times (panel (d)). We correctly determine $t^*=0.13 \, \mu s$, and obtain an MFPT estimation of $3.4 \, \mu s$, improving by an order of magnitude over standard iMetaD.

\subsection*{Chignolin mini-protein}

We close this paper with a more complex example -- the unfolding of chignolin in explicit water (simulations of 5889 atoms). Chignolin is a mini-protein composed of 10 amino acids\cite{lindorff-larsen_how_2011}, previously used to benchmark enhanced sampling methods\cite{miao_gaussian_2015,shaffer_enhanced_2016,ray_rare_2022,wang_efficient_2022,blumer2024combining}. A linear combination of six interatomic contacts, optimized via harmonic linear discriminant analysis (HLDA) by Mendels et. al.\cite{mendels_folding_2018}, serves as a good CV.
The radius of gyration (Rg) and the C-alpha root-mean-square deviation from a folded configuration (RMSD) serve as examples of suboptimal CVs. All simulation details are provided in the Methods section.

\begin{figure}[H]
  \includegraphics[width=\linewidth]{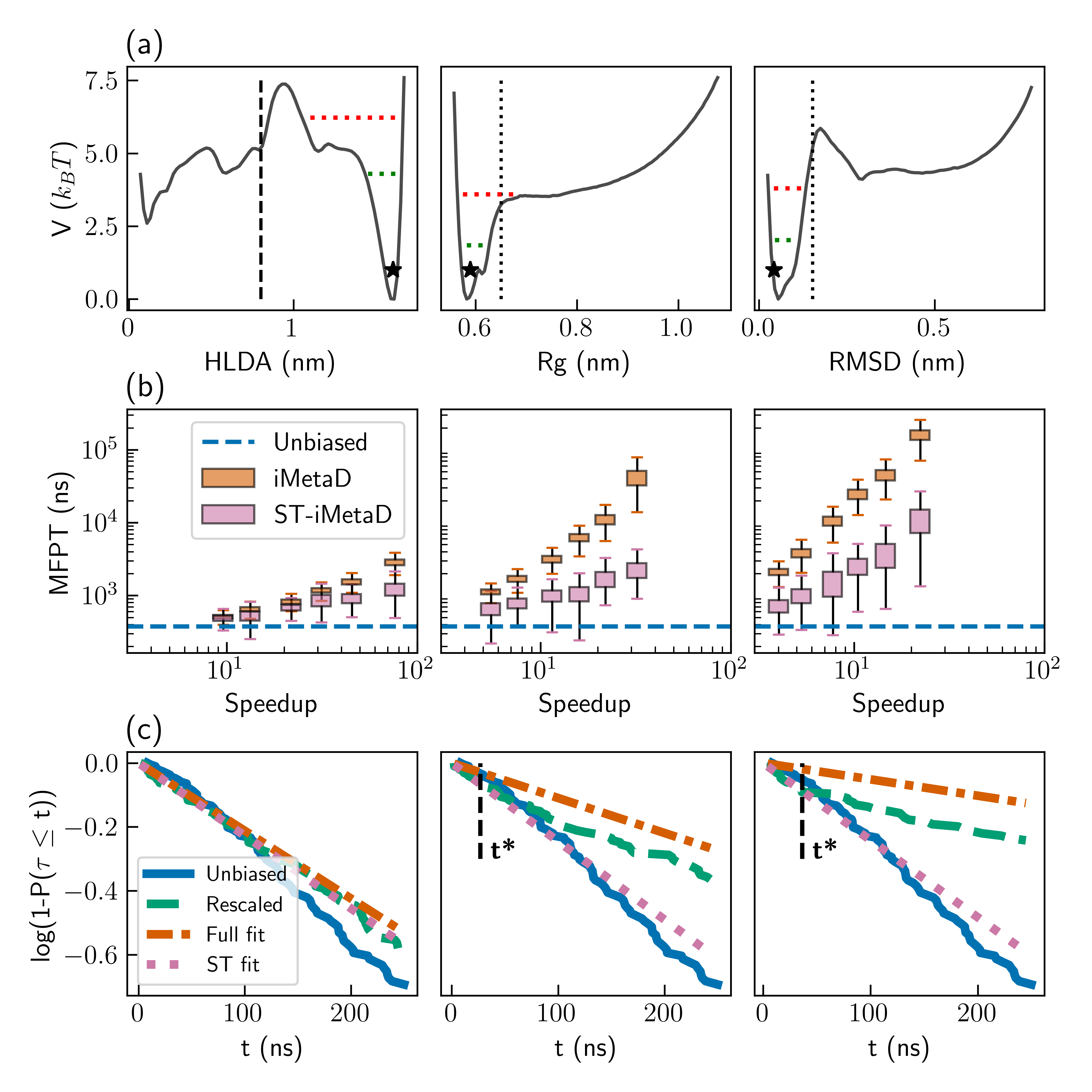}
  \caption{Results for simulation of solvated chignolin using the HLDA-based CV (left), the radius of gyration (middle), and the C-alpha RMSD (right). (a) Free-energy surfaces. The vertical dashed line marks the first-passage criterion, and the dotted vertical lines mark the average radius of gyration and C-alpha RMSD at first-passage events. The horizontal green and red lines highlight average maximal bias heights at $t^*$ and $10 t^*$, respectively. The black stars mark the values of the CVs at the initial folded configuration.
  (b) Estimated MFPTs obtained with standard iMetaD (orange) and ST-iMetaD (pink). The boxes show the range between the first and third quartiles and the whiskers show extreme values within 1.5 IQR below and above these quartiles. The blue dashed lines show the unbiased MFPT.
  (c) Survival functions for unbiased FTPs (blue solid lines), rescaled FPTs of simulations with a bias deposition rate of $1 \, ns^{-1}$(dashed green lines), exponential fits to the rescaled CDF in the entire range (orange dash-dotted lines), and linear fits to the survival function at $t \le t^*$ (pink dotted lines). The black dashed lines show the estimated $t^*$.} 
  \label{fig:chignolin}
\end{figure}

Figure \ref{fig:chignolin}(a) gives the free-energy surfaces (FES) along all CVs, obtained from umbrella sampling simulations (see Methods section for details). The values of the CVs at the initial folded configuration are marked with black stars.
We define the first-passage criterion as reaching a value $<0.8 \, nm$ for the HLDA-based CV (dashed black line). This process has an estimated MFPT of $\sim 376 \, ns$ (see SI for details). We note that the dynamics for reaching a stable unfolded state (HLDA$<0.2 \, nm$) leads to a MFPT that is longer by an order of magnitude~\cite{ray_rare_2022}. However, since the underlying assumptions of iMetaD are valid for escaping a single energy well, we limit our discussion to overcoming the first energy barrier along the HLDA-based CV. In addition, we note that the suboptimal CVs do not have a second minima in the FES for the stable unfolded state. Moreover, they do not fully distinguish between the unfolded and folded states. Therefore, even when biasing the suboptimal CVs, we use the value of the HLDA-based CV to determine the FPTs. The dotted lines in the middle and right panels of Figure \ref{fig:chignolin}(a), show the average values of those CVs when the first-passage criterion is fulfilled.

Figure \ref{fig:chignolin}(b) shows the estimated MFPT as a function of speedup, using the different CVs, with bias deposition rates in the range of $1$ to $50\,ns^{-1}$. We observe similar trends as in the previous examples.
In all cases, we find that ST-iMetaD leads to a better tradeoff between speedup and accuracy. Most notably, it is able to predict the MFPT rather successfully even for deposition rates where the standard approach leads to large errors. 

As with the former examples, we verify our assumptions by plotting the survival functions for the slowest bias deposition rate, $1\,ns^{-1}$ (Figure \ref{fig:chignolin}(c)). For all CVs, the rescaled and unbiased results match at short times. The suboptimal CVs
are associated with short $t^*$, while the value for the good CV is out of the scope of the plot ($425 \, ns$).

As a final test, we estimate the average maximum bias deposited up to $t^*$ and present it as a green horizontal dotted line in Figure \ref{fig:chignolin}(a). We find that it is lower than the barrier for both suboptimal and good CVs. This analysis provides insight into the onset of bias over-deposition. This is seen by looking at the average bias deposited at $10 t^*$, marked by horizontal dotted red lines in Figure \ref{fig:chignolin}(a), which are much closer to the barrier. This confirms that our procedure identifies the right $t^*$ within order of magnitude.

\section*{Conclusions}

To summarize, we present ST-iMetaD -- an improved inference scheme for iMetaD simulations. We find that the rescaled FPT distribution provides the correct short-time statistics, even for high bias deposition rates and suboptimal CVs. By focusing on these timescales, the time-independent rate of Poisson processes can be estimated reliably, resulting in a better tradeoff between speedup and accuracy in predicting the unbiased MFPTs.

The benefits of ST-iMetaD are demonstrated for the Wolfe-Quapp potential and two molecular systems, an isolated alanine dipeptide molecule and chignolin in explicit water. It reduces the prediction errors by orders of magnitude, especially for simulations with frequent bias deposition or suboptimal CVs. As a result, our method significantly extends the range of applicability of iMetaD, though it will eventually also break for unrealistically high deposition rates or exceedingly bad CVs. The ST-iMetaD scheme can be applied in post-processing of existing iMetaD data, with no additional cost in comparison to the standard approach, leading to an improved accuracy. Furthermore, the inference scheme is not limited to iMetaD, and could be applied to any enhanced sampling approach based on the iMetaD rescaling scheme, such as OPES-flooding~\cite{ray_rare_2022} or Variational Flooding~\cite{mccarty_variationally_2015,doi:10.1021/acs.jpclett.6b02852}.

\section*{Methods}


\subsection*{The Wolfe-Quapp potential}

Simulations of the Wolfe-Quapp potential were performed in the Large-scale Atomic/Molecular Massively Parallel Simulator (LAMMPS)~\cite{LAMMPS}. We followed the motion of a single particle with mass $m=40\, a.u.$ The simulations were carried out in the canonical (NVT) ensemble at a temperature of $300\, K$, using a Langevin thermostat~\cite{schneider_molecular-dynamics_1978}. The intergration timestep was $1 \, fs$ and the friction coefficient was $0.01 \, fs^{-1}$. 
MetaD was implemented using PLUMED 2.7.1.~\cite{bonomi_plumed_2009,tribello_plumed_2014,Bonomi2019} We used a bias height of $0.5 \, k_B T$, bias factor of $5$, bias width of $\sigma = 0.1\, nm$ and grid spacing of $0.01 \, nm$.

\begin{figure}
  \includegraphics[width=0.5\linewidth]{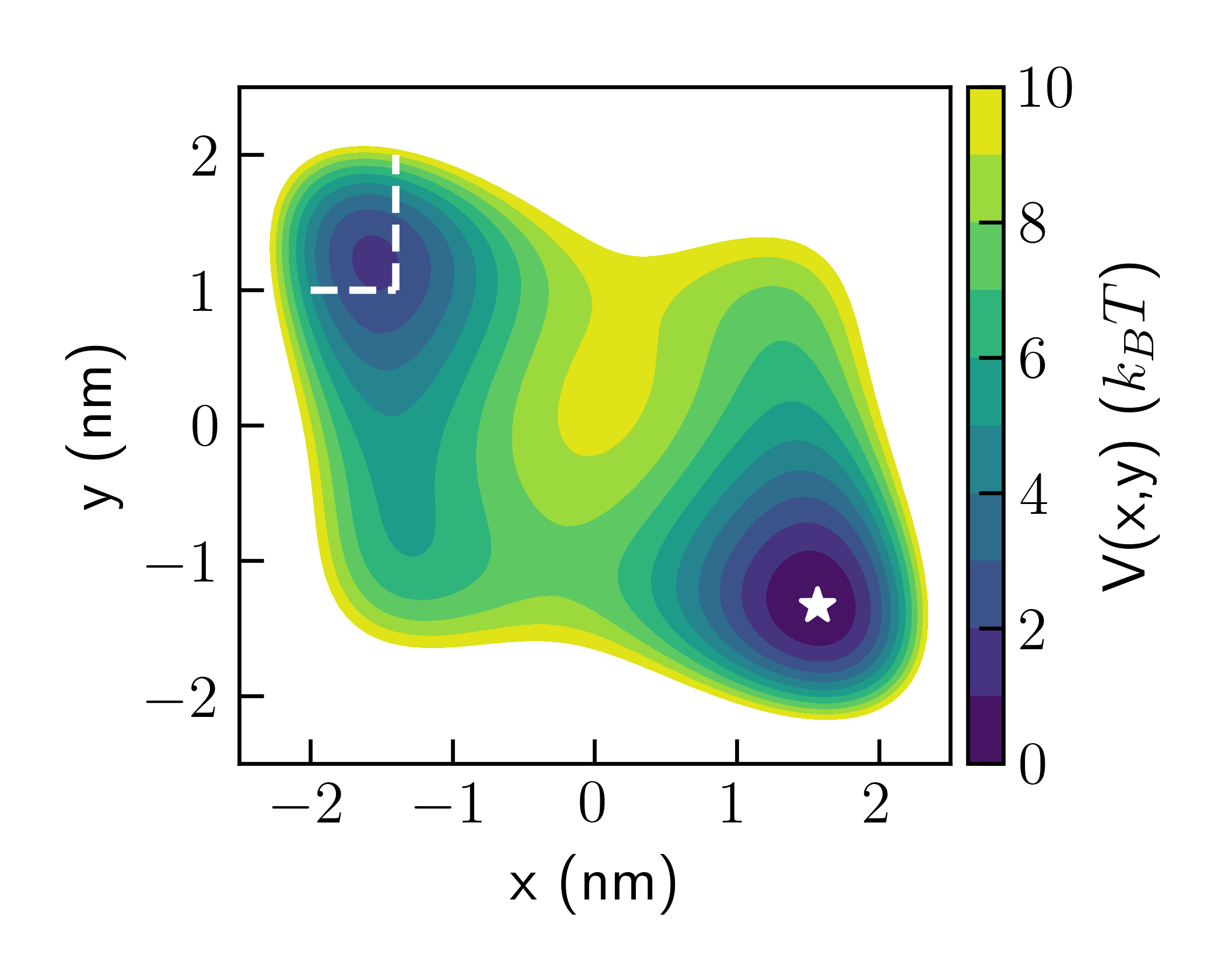}
  \caption{The Wolfe-Quapp potential. The initial position is marked with a star, and the target basin is marked with dashed lines.}
  \label{fig:WQ}
\end{figure}

The external potential was implemented in the LAMMPS input files. Its structure is as described in previous publications~\cite{invernizzi_making_2019,ray_rare_2022}, and is shown in Figure \ref{fig:WQ}. The exact form used is given in Eq. \ref{eqn:WQ}, with the distance given in units of $nm$ and the energy in units of $1\,k_B T$:

\begin{equation}
 V(x,y) = x^4+y^4-4x^2-2y^2+2xy+0.1x+0.8y
 \label{eqn:WQ}
\end{equation}

Simulations were initiated from the global minimum $(x=1.564,y=-1.334)\, nm$, marked with a star in Figure \ref{fig:WQ}, with velocities sampled from the Maxwell Boltzmann distribution. All trajectories were stopped using the COMMITTOR command in PLUMED when reaching the second local minimum, defined as $x<-1.4 \, nm \land y>1.0 \, nm$, denoted by dashed lines in Figure \ref{fig:WQ}.

\subsection*{Alanine dipeptide}

Simulations of alanine dipeptide in vacuum were performed in GROMACS 2019.6~\cite{abraham_gromacs_2015}. We used input files by Bonomi and Bussi~\cite{noauthor_plumed_nodate}, implementing the AMBER99SB force field (FF). Simulations were performed in the NVT ensemble at a temperature of $300\, K$, using a stochastic velocity rescaling thermostat~\cite{bussi_canonical_2007}. The integration time step was $2 \, fs$. MetaD was implemented once again using PLUMED. We used a bias height of $0.5 \, k_B T$, bias factor of $5$, bias width of $\sigma = 0.25\, \text{rad}$ and grid spacing of $0.01 \, \text{rad}$. All trajectories were initiated in a fixed position at the $C_{7eq}$ conformer and stopped using PLUMED when reaching $0.5<\phi<1.5 \, \text{rad}$. The stopping criterion was checked every $1\, ps$.

\subsection*{Chignolin}

Simulations of chignolin in water were performed using the same software as those of alanine dipeptide. We used input files by Ray et. al.~\cite{ray_rare_2022}, available at PLUMED-NEST, the public repository of the PLUMED consortium~\cite{Bonomi2019}, as plumID:22.031. We used the CHARMM22* FF~\cite{piana_how_2011} for the protein and the CHARMM TIP3P FF~\cite{mackerell_all-atom_1998} for water. The thermodynamic ensemble, thermostat and integration timestep were the same as employed for alanine dipeptide, but the temperature was higher, $340\, K$. 

We used a bias height of $0.5 \, k_B T$, bias factor of $5$ and grid spacing of $0.001\, nm$ for all CVs. 
The bias width was $0.022,\,0.005$ and $0.006\, nm$ for the HLDA, Rg and RMSD-based CVs, respectively. 
All trajectories were initiated from a fixed position and stopped using PLUMED when reaching $s  < 0.8 \, nm$, with $s$ being the HLDA-based CV. This stopping criterion was checked every $1\, ps$.

To construct the FES featured in Figure 4(a), we performed 32, $100 \, ns$ long umbrella sampling simulations~\cite{torrie_nonphysical_1977,kastner_umbrella_2011}for each CV, with harmonic constraints centered at $s_{min}+i\Delta_s$, with $i$ going from $0$ to $31$. We used $s_{min}=0.5,6,0.25\, \AA$ and $\Delta_s =0.5, 0.167,0.25\, \AA$ for the HLDA, Rg and RMSD-based CVs, respectively. 
The harmonic constant was $k=3 \, k_B T\,$\AA$^{-2}$ for all CVs. The value of the CV was saved every $1\,ps$, and the FES was constructed through the weighted histogram analysis method (WHAM), using the implementation of Grossfield~\cite{wham}.

\section*{Data Availability}

Example input files, source data, and an example analysis script to perform ST-iMetaD are available in the GitHub repository: \href{https://github.com/OfirBlumer/ST-iMetaD}{\url{https://github.com/OfirBlumer/ST-iMetaD}}.

\section*{Associated Content}

Supporting Information available. Details on the committor analysis and on the MFPT estimation, supplementary figures showing the sensitivity of the results to the bootstrapping batch size and a comparison to the KTR method.

\begin{acknowledgement}
Barak Hirshberg acknowledges support by the Israel Science Foundation (grants No. 1037/22 and 1312/22) and the Pazy Foundation of the IAEC-UPBC (grant No. 415-2023). This project has received funding from the European Research Council (ERC) under the European Union’s Horizon 2020 research and innovation program (grant agreement No. 947731).
\end{acknowledgement}

\bibliography{achemso-demo}

\end{document}


\section{Committor analysis}

Here we report the details of the committor analysis performed to estimate the quality of different collective variables (CVs) for the Wolfe-Quapp potential. Our procedure is based on that described by Peters~\cite{PETERS2017539}. We defined two states, $A$ at $x>1.4 \, nm$, $y<-1.0 \, nm$ and $B$ at $x<-1.4 \, nm$, $y>1.0 \, nm$. For each examined CV $s(x,y)$, we first performed a $10 \, ns$ long simulation initiated at $(x=0,y=0) \, nm$, restricted to $s \approx 0 \, nm$ by a bias potential $V(s) = 80s^2$, with $V(s)$ in units of $1 \, k_B T$ and $s$ in $nm$. A thousand uncorrelated configurations were randomly sampled from the restricted trajectory, to serve as new initial configurations. Next, we sampled 100 trajectories for each initial configuration and obtained the fraction of trajectories that reached the target state $B$ before state $A$, $p_B$. 

\begin{figure}
  \includegraphics[width=\linewidth]{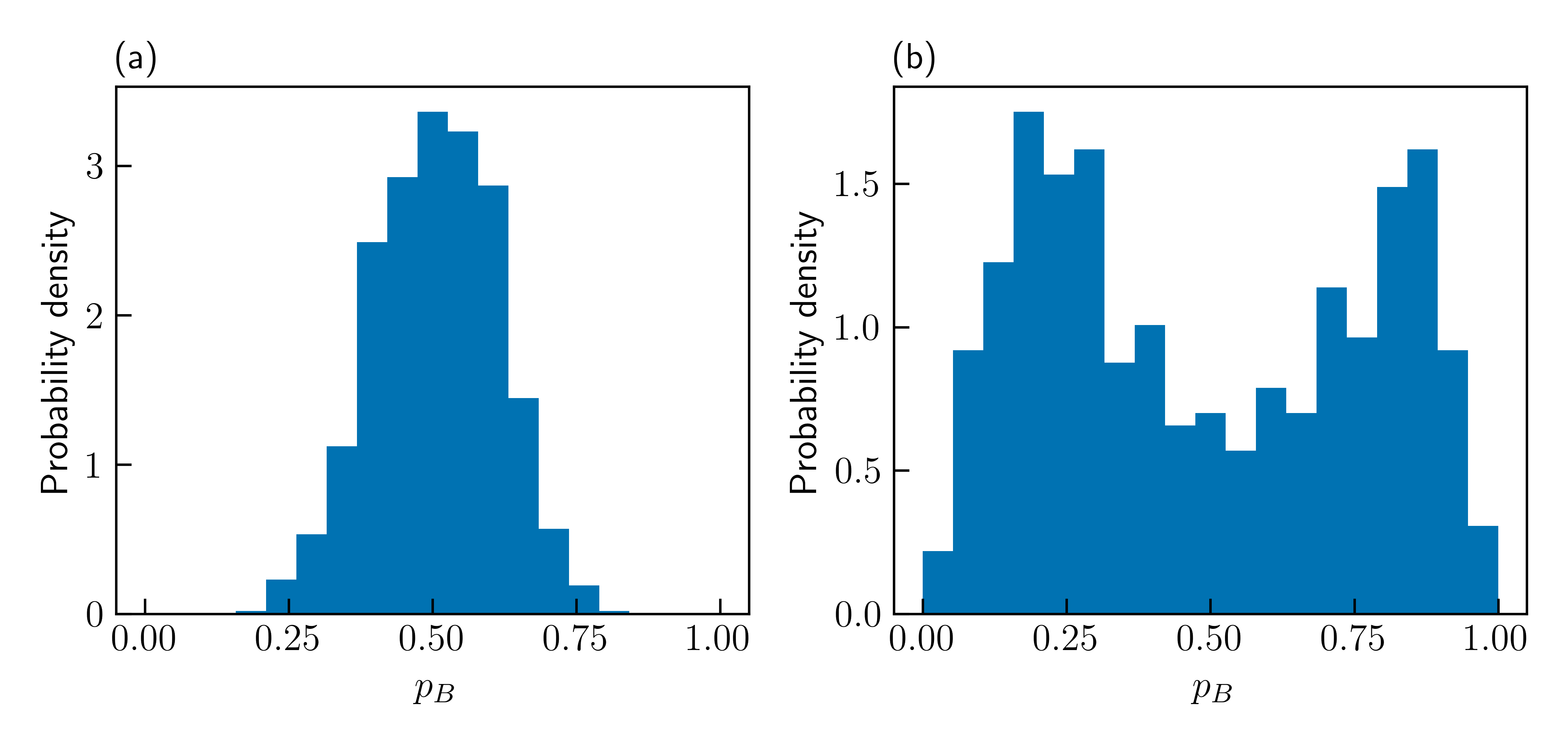}
  \caption{Histograms of $p_B$ for (a) the good CV and (b) a poor CV.}
  \label{fig:committor}
\end{figure}

Finally, we examined the histogram of $p_B$ for all tested CVs. The CV showing the narrowest distribution around $p_B=0.5$ best approximates the true committor and is regarded in the main text as the good CV. It is given by $s(x,y) = \cos \left(\frac{\pi}{9}\right) x - \sin\left(\frac{\pi}{9}\right) y$.
Figure \ref{fig:committor} shows the histograms of $p_B$ for this CV and for $s(x,y)=x$, which is a poor CV, for comparison.

\section{Mean first-passage time estimation}

Here we explain how the unbiased mean first-passage time (MFPT) values of the different processes were evaluated. For all systems, we first performed standard molecular dynamics simulations, that were stopped when the first-passage criterion was fulfilled. For the Wolfe-Quapp potential, all trajectories exhibited a first-passage. The MFPT was simply taken as the average first-passage time (FPT) of all trajectories. However, for both molecular systems, $\sim 7\%$ of the simulations did not show a first-passage event. Therefore, taking the MFPT of all trajectories would underestimate the true value.

Instead, we used the assumption that the FPT distribution is exponential. We made a linear fit to the logarithm of the survival function, and evaluated the true MFPT as $-k^{-1}$, with $k$ being the slope of the fit. For both systems, we obtained high coefficient of determination values, $R^2 > 0.995$, confirming that the underlying distributions are exponential. 

\section{Sensitivity of $t^*$ to batch size}

Here, we plot the values of $t^*$ as a function of the bootstrapping batch size, for the systems shown in Figure 3 of the manuscript.

\begin{figure}
  \includegraphics[width=\linewidth]{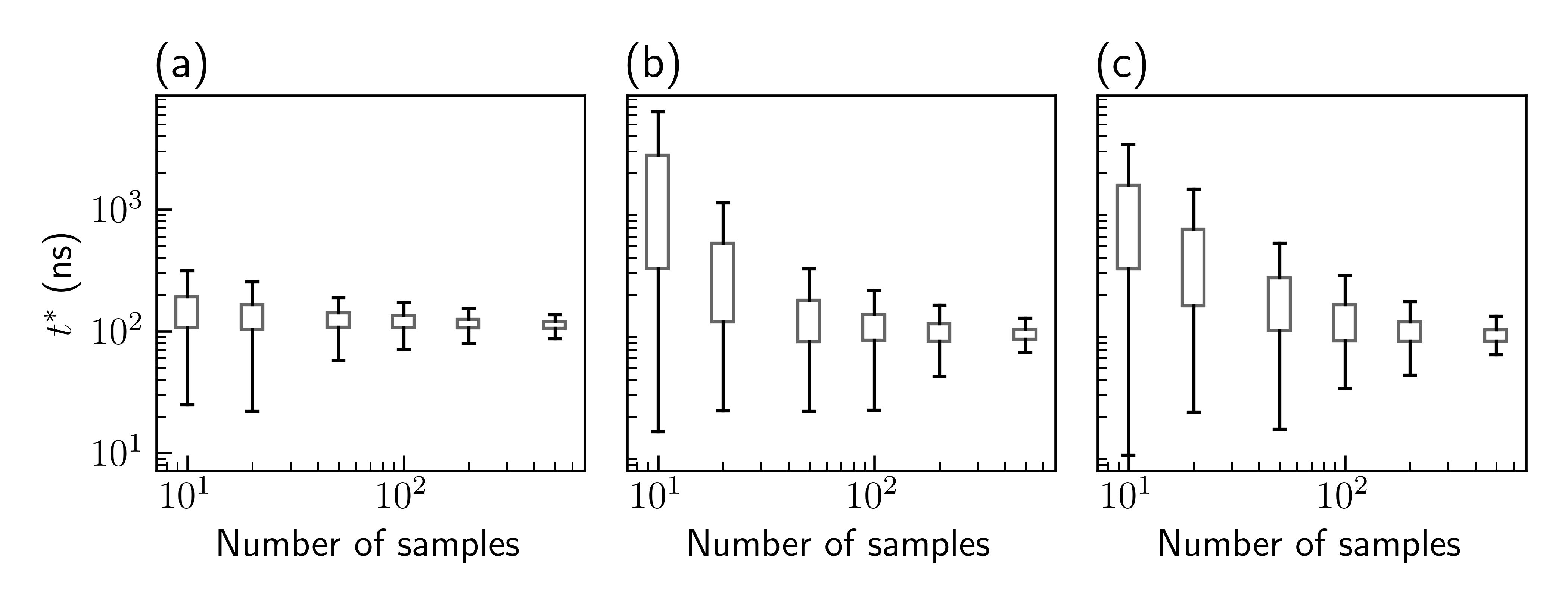}
  \caption{$t^*$ as a function of the number of samples in each bootstrapping batch for simulations using (a) a good CV and bias deposition rate of $10 \, ns^{-1}$, (b) a good CV and bias deposition rate of $1000 \, ns^{-1}$, and (c) a suboptimal CV and bias deposition rate of $200 \, ns^{-1}$. The boxes show the range between the first and third quartiles (interquartile range, IQR) and the whiskers show extreme values within 1.5 IQR below and above these quartiles.}
  \label{fig:tstar}
\end{figure}

\section{Sensitivity of estimated MFPT to batch size}

Here, we provide plots equivalent to Figure 2, Figure 4(a), Figure 4(b), and Figure 5(b) of the manuscript, with smaller bootstrapping batch sizes $N$.

\begin{figure}
  \includegraphics[width=\linewidth]{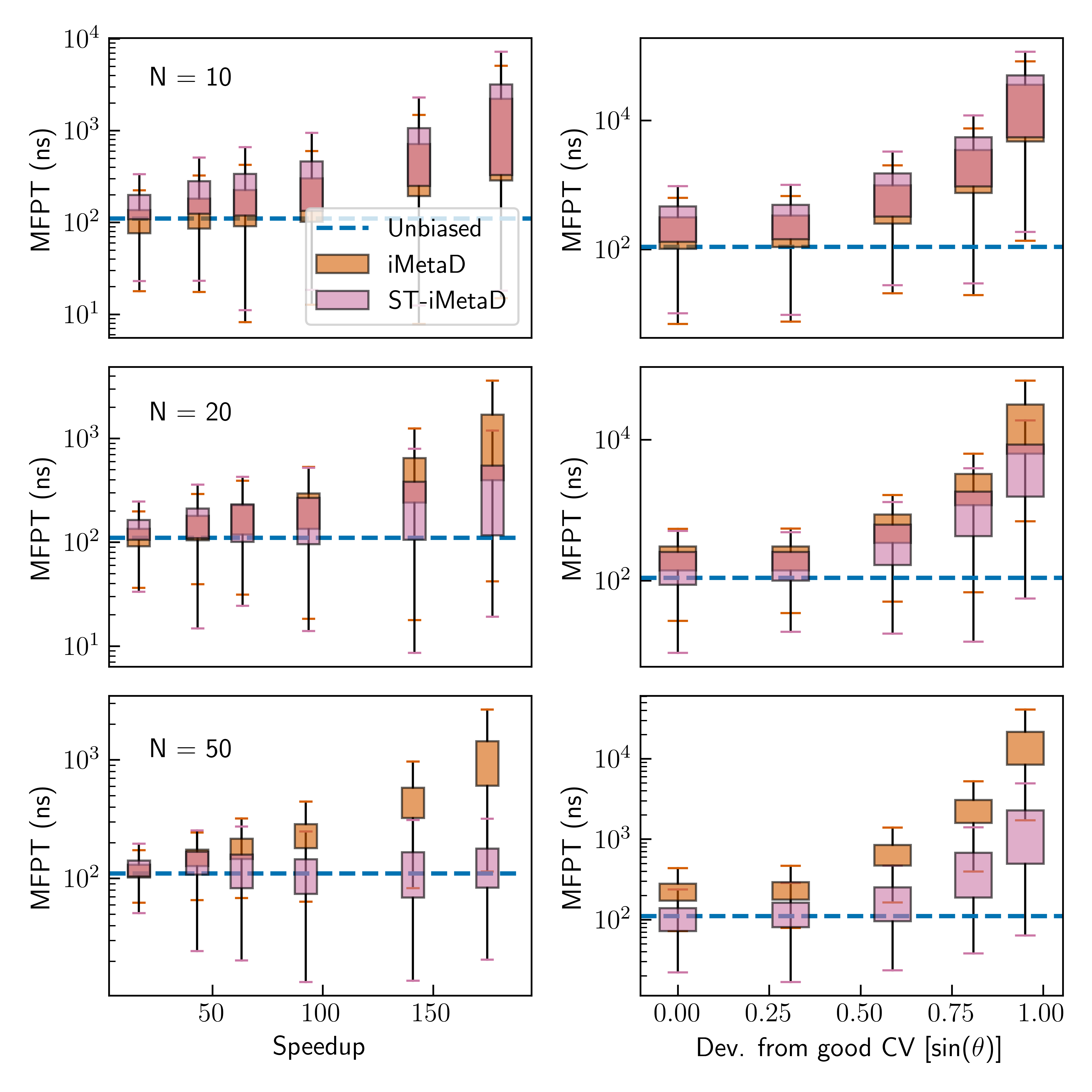}
  \caption{Results for the Wolfe-Quapp potential with sets of 10 (upper row), 20 (middle row), or 50 (bottom row) trajectories in each bootstrapping batch, using standard iMetaD (orange) or ST-iMetaD (pink). Left: Estimated MFPT as a function of speedup for simulations using a good CV and different bias deposition rates from $10$ to $1000\, ns^{-1}$. Right: Estimated MFPTs for a bias deposition rate of $200 \, ns^{-1}$, and different choices of CV. The boxes show the range between the first and third quartiles and the whiskers show extreme values within 1.5 IQR below and above these quartiles. The blue lines show the unbiased MFPT.}
  \label{fig:WQdifferentN}
\end{figure}

\begin{figure}[H]
  \includegraphics[width=\linewidth]{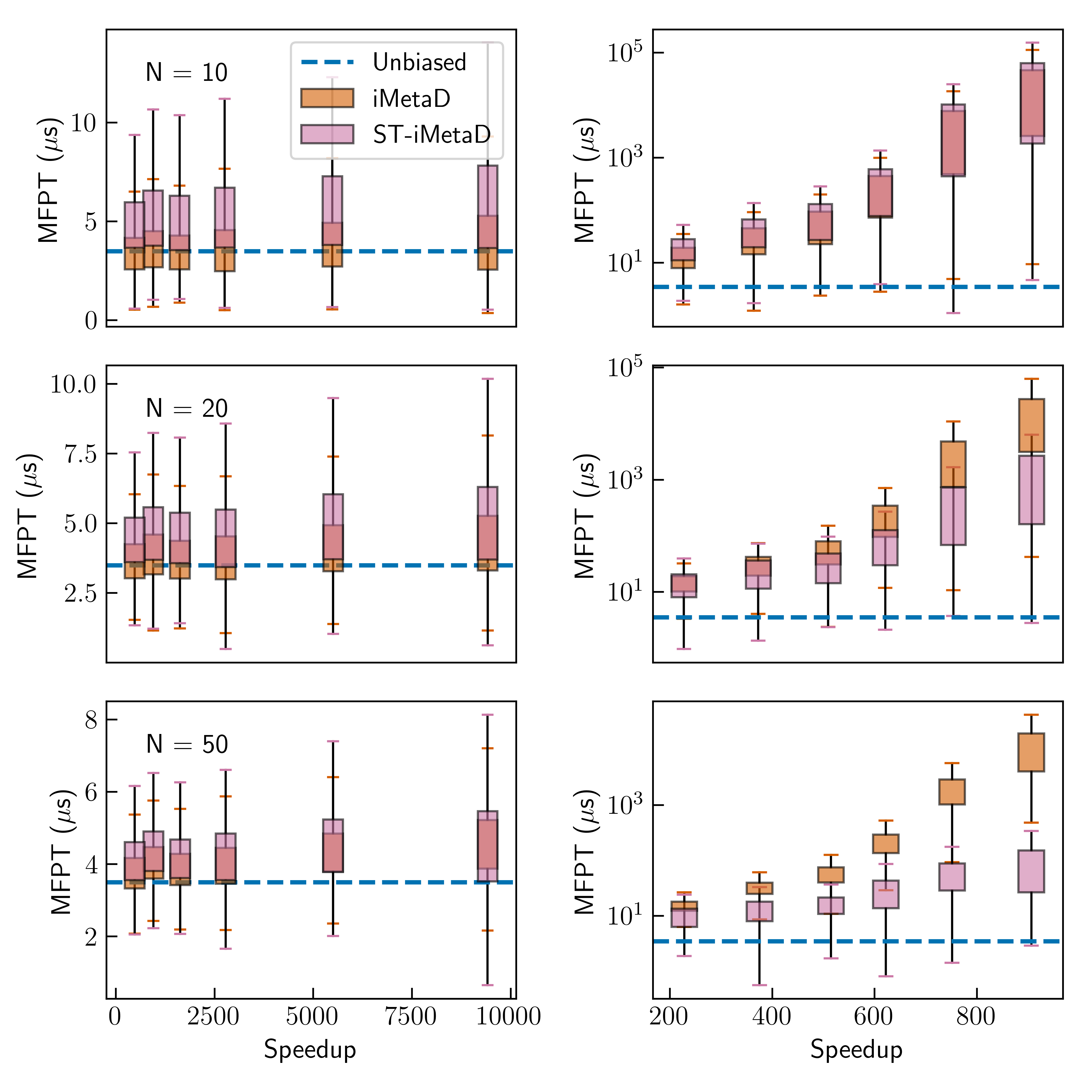}
  \caption{Results for alanine dipeptide with sets of 10 (upper row), 20 (middle row), or 50 (bottom row) trajectories in each bootstrapping batch, using the $\phi$ angle (left) or $\psi$ angle (right) as CV. The blue lines show the unbiased MFPT while the boxes give estimations through standard iMetaD (orange) or ST-iMetaD (pink). The boxes show the range between the first and third quartiles and the whiskers show extreme values within 1.5 IQR below and above these quartiles.}
  \label{fig:WQdifferentN}
\end{figure}

\begin{figure}[H]
  \includegraphics[width=\linewidth]{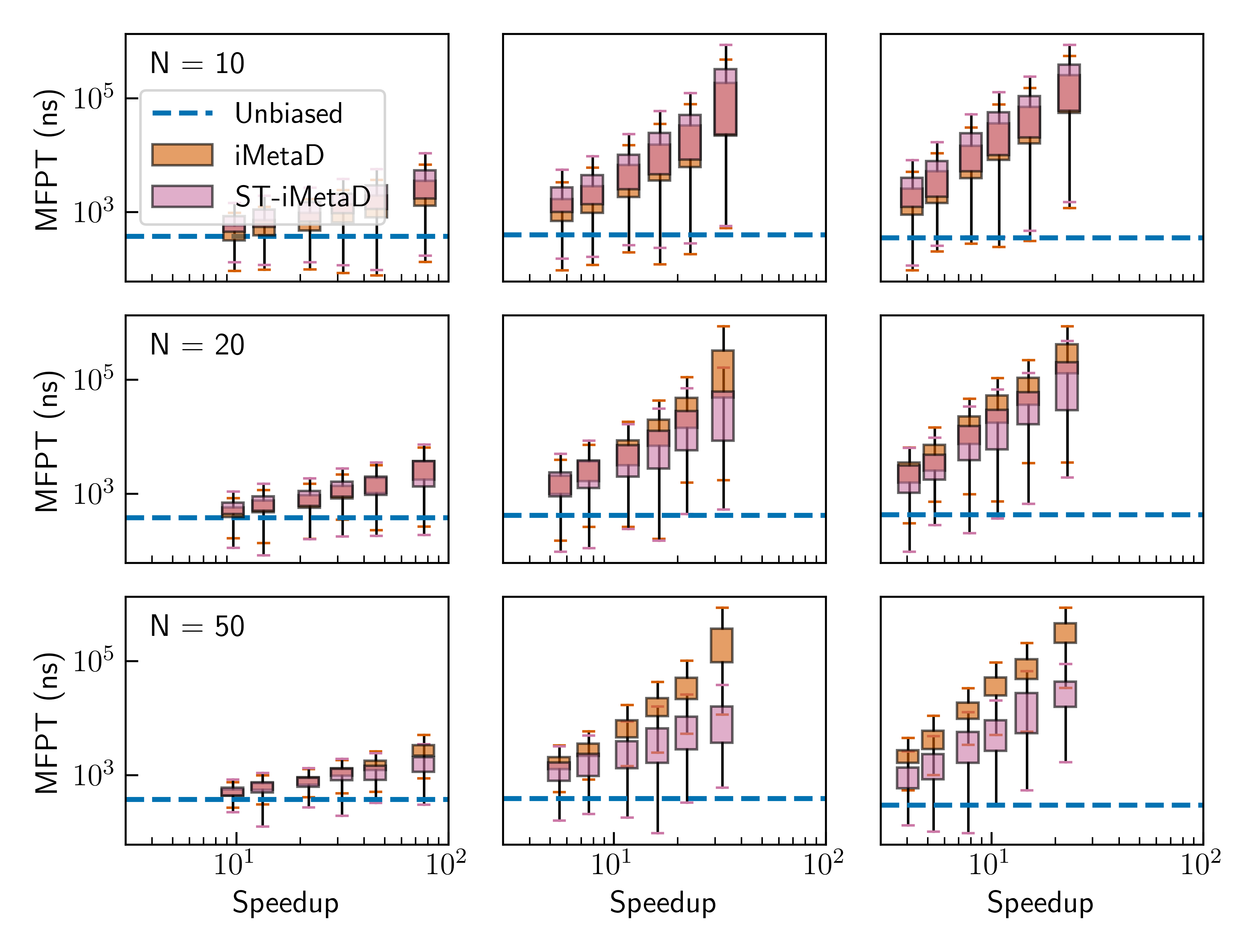}
  \caption{Results for chignolin unfolding with sets of 10 (upper row), 20 (middle row), or 50 (bottom row) trajectories in each bootstrapping batch, using CVs based on either HLDA (left column), Rg (center column), or RMSD (right column). The blue lines show the unbiased MFPT and the boxes give estimations through standard iMetaD (orange) or ST-iMetaD (pink). The boxes show the range between the first and third quartiles and the whiskers show extreme values within 1.5 IQR below and above these quartiles.}
  \label{fig:CHIGdifferentN}
\end{figure}

\section{Comparison with Kramers Time-Dependent Rate}

Here, we compare between ST-iMetaD and Kramers Time-Dependent Rate (KTR) for the three systems presented in the manuscript.

KTR is a recently developed kinetics inference scheme from biased simulations~\cite{palacio-rodriguez_transition_2022}.
Relying on Kramers theory~\cite{KRAMERS1940284,PETERS2017435}, it assumes that the kinetic rate of the biased process at time $t$ is given by $k(t) = k_0 e^{\beta \gamma V_{MB}(t)}$,
with $k_0$ being the unbiased kinetic rate, $\beta$ the inverse temperature, and $V_{MB}(t)$ the average maximum height of the biasing potential. $k_0$ and $\gamma \in [0,1]$ are obtained from a maximum likelihood fitting to the empirical survival function.

We reproduce Figure 2, Figure 4(a), Figure 4(b), and Figure 5(b) of the manuscript, adding the MFPT estimations through KTR in red. We used open-source code provided by the authors of the method~\cite{KTR}.
We find that KTR underestimates the MFPT for simulations with high bias deposition rates or suboptimal CVs, as opposed to iMetaD which overestimates it. ST-iMetaD provides better predictions than KTR for the Wolfe-Quapp potential and alanine dipeptide. For Chignolin, in most CVs and bias deposition rates, ST-iMetaD provides similar accuracy to KTR, expect for the RMSD-based CV. This case suffers the most from bias over-deposition, that overestimates the MFPT, which possibly results in a fortuitous cancellation of errors for KTR.

\begin{figure}[H]
  \includegraphics[width=\linewidth]{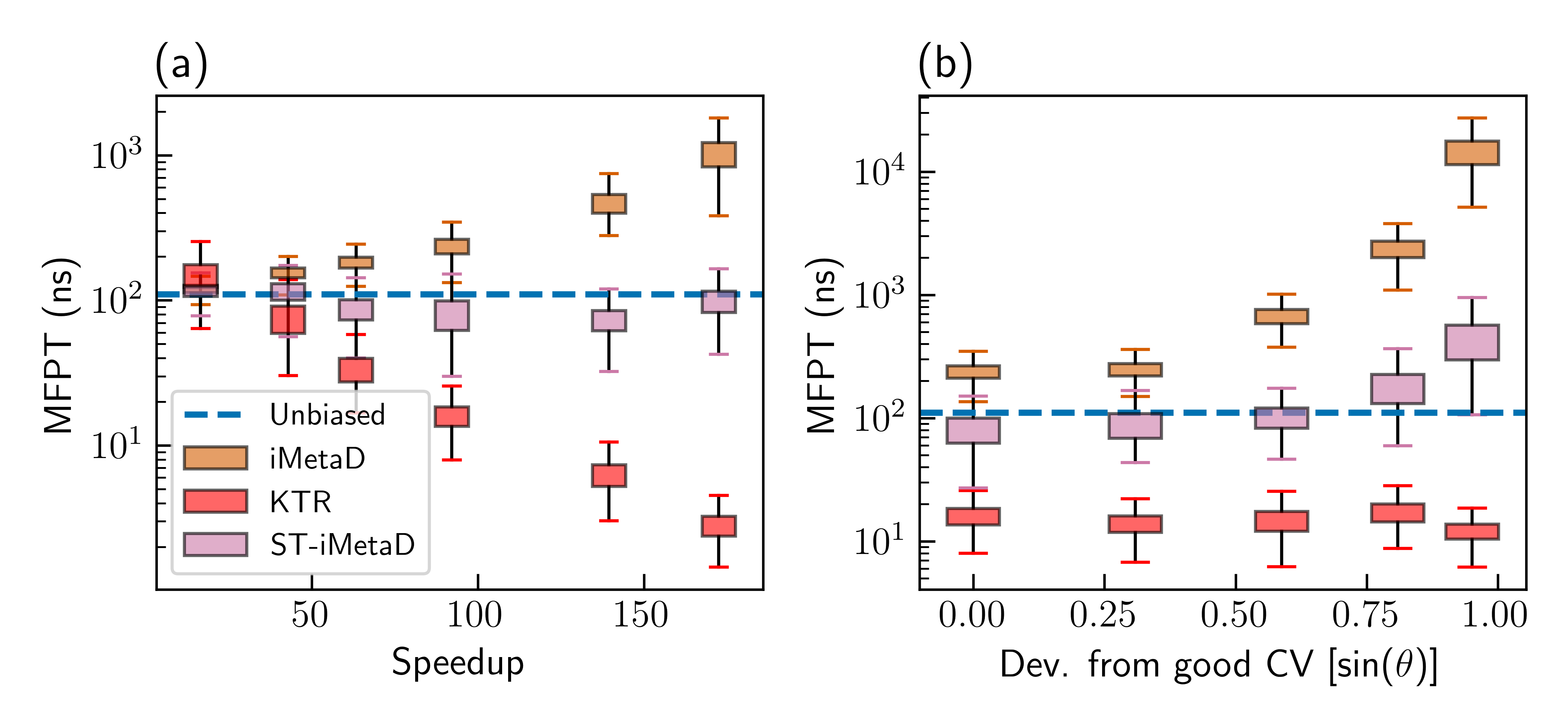}
  \caption{Results for the Wolfe-Quapp potential. (a) Estimated MFPT as a function of speedup for simulations using a good CV and different bias deposition rates from $10$ to $1000\, ns^{-1}$. (b) Estimated MFPTs for a bias deposition rate of $200 \, ns^{-1}$, and different choices of CV. We used either standard iMetaD (orange), KTR (red), or ST-iMetaD (pink). The boxes show the range between the first and third quartiles and the whiskers show extreme values within 1.5 IQR below and above these quartiles. The blue lines show the unbiased MFPT.}
  \label{fig:WQKTR}
\end{figure}

\begin{figure}[H]
  \includegraphics[width=\linewidth]{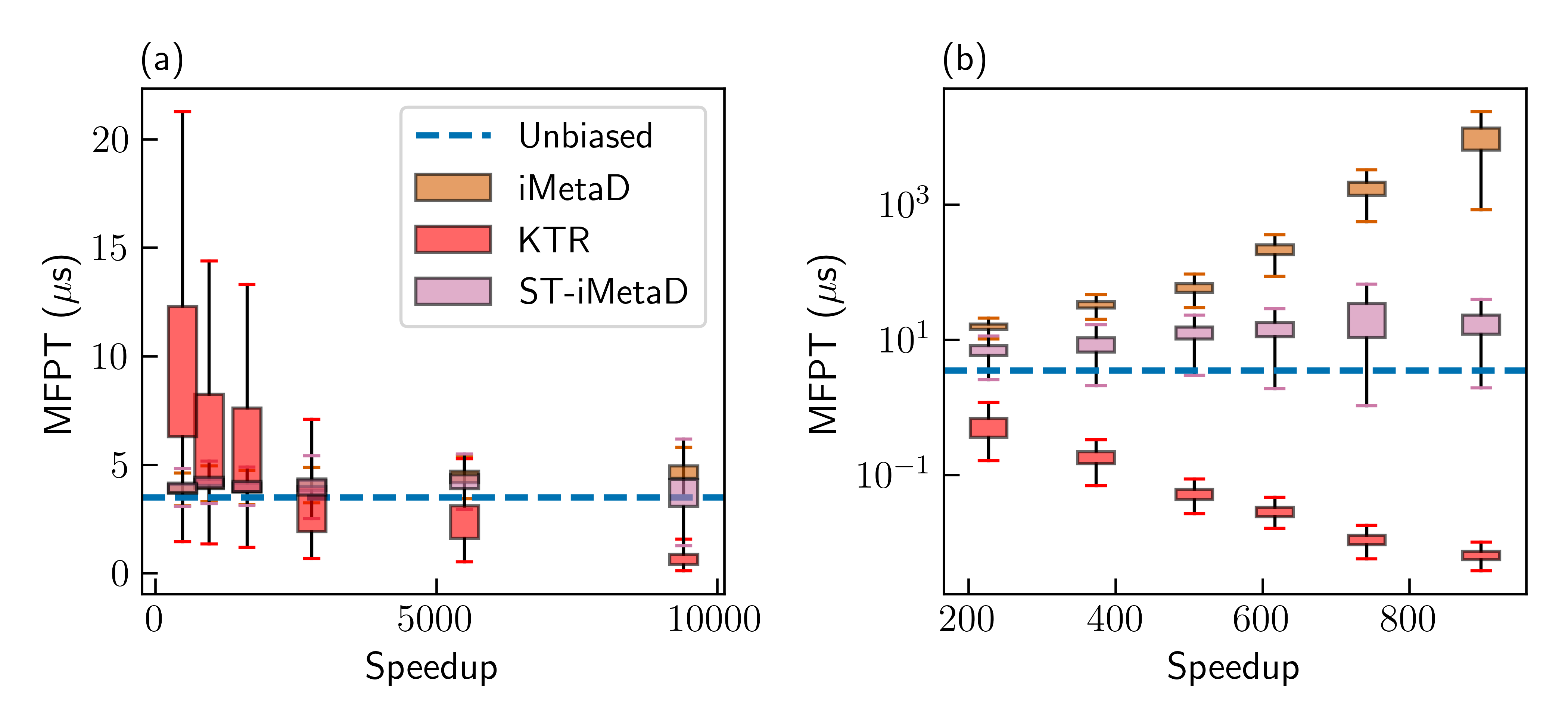}
  \caption{Results for alanine dipeptide using (a) the $\phi$ angle or (b) the $\psi$ angle as CV. We used either standard iMetaD (orange), KTR (red), or ST-iMetaD (pink). The boxes show the range between the first and third quartiles and the whiskers show extreme values within 1.5 IQR below and above these quartiles. The blue lines show the unbiased MFPT.}
  \label{fig:ala2KTR}
\end{figure}

\begin{figure}[H]
  \includegraphics[width=\linewidth]{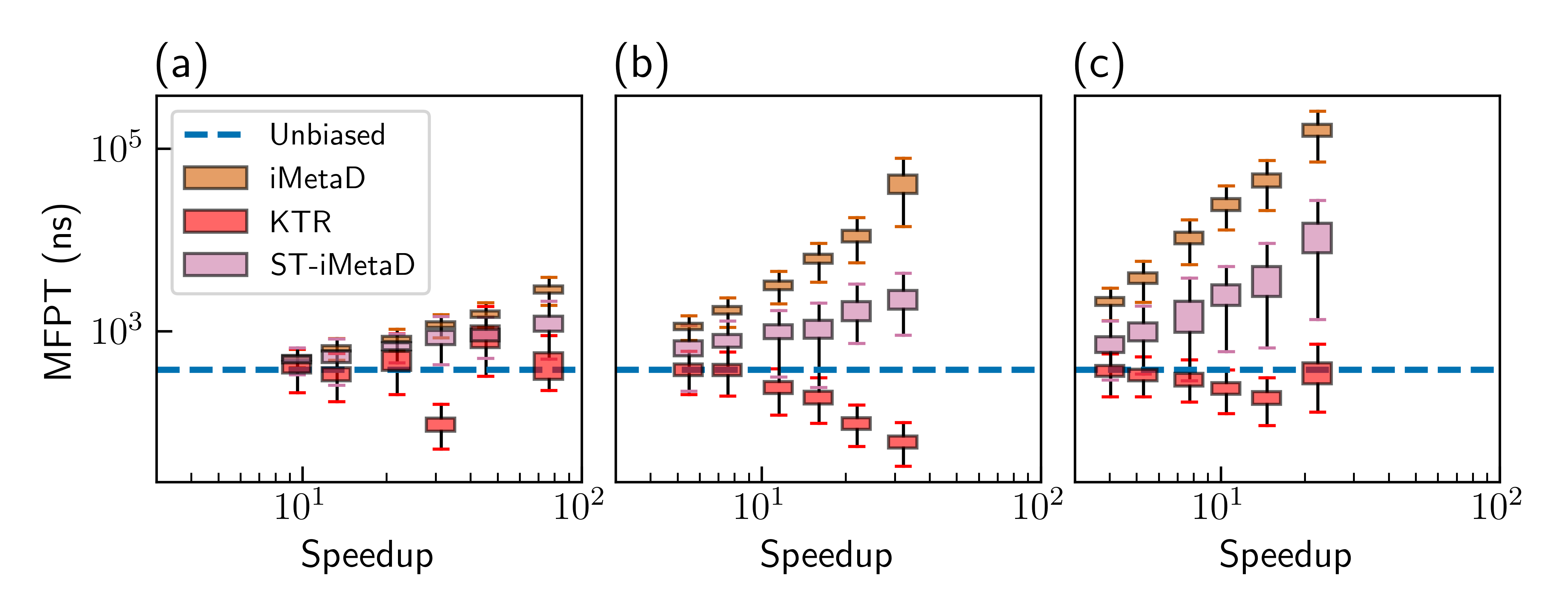}
  \caption{Results for chignolin using (a) the HLDA, (b) the Rg, or (c) the RMSD-based CVs. We used either standard iMetaD (orange), KTR (red), or ST-iMetaD (pink). The boxes show the range between the first and third quartiles and the whiskers show extreme values within 1.5 IQR below and above these quartiles. The blue lines show the unbiased MFPT.}
  \label{fig:chignolinKTR}
\end{figure}

\bibliography{bibsi}